\documentstyle[11pt,paspconf]{article}
\input epsf

\begin{document}

{\it Invited talk at IAGUSP Workshop on Young Galaxies and QSO Absorbers}

\title{Cosmic Minivoids in the Intergalactic Medium}

\author{Avery Meiksin}
\affil{Department of Astronomy and Astrophysics, University of Chicago,
5640 S. Ellis Ave., Chicago, IL\ 60637;\ meiksin@oddjob.uchicago.edu}

\begin{abstract}
The Gunn-Peterson effect, absorption of Ly$\alpha$ photons by a homogeneous
component of the intergalactic medium (IGM), potentially provides a test of
Big Bang Nucleosynthesis (BBN). With a lower limit on the UV radiation
field estimated from the contribution due to QSOs, a measurement of the
Ly$\alpha$ opacity of the intergalactic medium would permit the
derivation of a lower bound to the baryonic density of the universe.
The effect, however, has continually eluded a convincing detection,
both in {H$\;${\small\rm I\relax}} and {He$\;${\small\rm II\relax}},
despite extensive searches. Recent cosmological hydrodynamical simulations
of structure formation in the intergalactic medium suggest an explanation
for its absence. In a Cold Dark Matter dominated cosmology, the fragmentation
of the baryons is nearly complete, leaving a negligible remnant to comprise
a smoothly distributed component. The fragmentation extends even into regions
that are underdense, where it gives rise to most of the
optically thin {H$\;${\small\rm I\relax}} systems and nearly all of the
{He$\;${\small\rm II\relax}} systems, both thin and saturated. The result is
a Ly$\alpha$ opacity from a smooth IGM that is suppressed by over two orders
of magnitude from the BBN value.
\end{abstract}

\section{Introduction}

Soon after the identification of the first QSO, Gunn \& Peterson
(1965) recognized that its spectrum could be used to place a stringent
limit on the density of neutral hydrogen in the intergalactic
medium (IGM). Photons blueward of the frequency of Ly$\alpha$
will be absorbed by neutral hydrogen en route to the observer in an
expanding homogeneous IGM while photons redward will be transmitted
(Field 1959). Hence, a step in the spectrum of a QSO is expected shortward of
its Ly$\alpha$ emission line. Gunn and Peterson found a step
corresponding to an opacity of $\sim1/2$, and a neutral comoving
{H$\;${\small\rm I\relax}} density of $n_{\rm
HI}\approx2\times10^{-12}\,{\rm cm^{-3}}$. This is
only a tiny fraction of the density of hydrogen in galaxies
today. Their inference was that the IGM must be highly ionized. The
detection of absorption by a homogeneous ionized component, the
``Gunn-Peterson effect'', would have important implications for
cosmology. If the gas is photoionized, then the equation of
photoionization equilibrium may be solved for the baryon density of
the IGM, given an estimate of the ionizing radiation field.  A lower
bound on the radiation field may be placed by summing the contribution
due to QSOs, which would then permit a lower bound to be set on the
density of the IGM. This would be a direct test of Big Bang
Nucleosynthesis (BBN).

Subsequently, however, more detailed spectra revealed a host of
absorption lines shortward of Ly$\alpha$. These were identified by
Lynds (1971) as Ly$\alpha$ absorption by intervening discrete
absorbers, the Ly$\alpha$ forest. The original opacity measured by
Gunn and Peterson could be accounted for by the forest. Because the
forest is clumped, it is not possible to solve for the baryon density
of the IGM independently of a cloud model. Searches since then for
absorption by a homogeneous component of the IGM (e.g., Steidel \&
Sargent 1987), led to similar results: all the absorption could be
accounted for by the Ly$\alpha$ forest. The homogeneous unclustered
intergalactic medium expected from Big Bang nucleosynthesis has continued to
evade detection. Recent numerical hydrodynamical computations of structure
formation in Cold Dark Matter (CDM) dominated cosmologies suggest an
explanation. These simulations show that all the baryons undergo
condensation into discrete systems, even when the density of the
system is less than the cosmic mean. These systems give rise to
absorption lines that are optically thin at line center, and form in
regions a few megaparsecs across that are themselves underdense,
``cosmic minivoids.''

\section{Baryons Lost}

The Ly$\alpha$ opacity of homogeneously distributed intergalactic
hydrogen photoionized by an ambient metagalactic radiation field is

\begin{equation}
\tau_\alpha\simeq0.18h_{50}^{-1}T_4^{-0.75}(\Omega_Dh_{50}^2)^2
(1+z)^5(1+2q_0z)^{-1/2}\Gamma_{\rm HI,-12}^{-1},
\end{equation}
where $\Omega_D$ is the total baryonic density of the diffuse homogeneous
component in terms of the closure density, $T_4$ is the temperature of
the neutral hydrogen in units of $10^4\,{\rm K}$, $\Gamma_{\rm HI,-12}$
is the photoionization rate of the neutral hydrogen in units of
$10^{-12}\,{\rm s^{-1}}$, $q_0$ is the cosmological deceleration parameter, and
$h_{50}$ is the Hubble constant in units of $50\,{\rm km\,s^{-1}\,Mpc^{-1}}$.
In the redshift range $2<z<3$, the contribution of QSOs to the photoionizing
background gives $\Gamma_{\rm HI,-12}\approx1$, with an uncertainty of
perhaps a factor of 2--3 (Meiksin \& Madau 1993; Haardt \& Madau 1996). By
$z=4$ this value may decline by a factor of 2, and by a factor of 20 by
$z=5$, although the QSO counts become increasingly uncertain at these high
redshifts. The baryon density due to BBN is estimated to lie in the range
$0.04\la\Omega_bh_{50}^2\la0.08$ ($2\sigma$, Copi et al. 1995), consistent
(though just barely!) with recent $D/H$ measurements in intergalactic gas
clouds (Tytler et al. 1996; Rugers et al. 1996). The expected IGM opacity
is thus $\tau_\alpha\approx0.15$ at $z=3$ and 0.8 at $z=4$. These values
greatly exceed current ($2\sigma$) limits. Steidel \& Sargent (1987) find
$\tau_\alpha<0.08$ at $z\approx3$, while Giallongo et al. (1994) find
$\tau_\alpha<0.08$ at $z\approx4.3$, although perhaps a more conservative
upper limit is that of Jenkins \& Ostriker (1991) of $\tau_\alpha<0.6$ at
$z\approx4$.

The conflict between the expected and measured upper limits may be
alleviated by postulating radiation sources in addition to QSOs, like
stars or decaying particles, or if a substantial fraction of high
redshift QSOs were missed in optical surveys due to obscuration by
dust in intervening absorbers (Fall \& Pei 1993). An increase by a
factor of up to 10 would be consistent with the proximity effect
(Bajtlik et al. 1988), although smaller values derived from the
proximity effect may be favored (Espey 1993). Alternatively, Meiksin
\& Madau (1993) suggested that the conflict may indicate a very clumpy
IGM. They argued that a large fraction of the baryons may be contained
in the Ly$\alpha$ forest, given the observed numbers and size constraints
of the absorbers. Their estimate was  $0.002 <
\Omega_{\rm Ly\alpha}h_{50}<0.05$ for the density parameter of baryons in
the forest.

\section{Numerical Simulations}

It has recently become possible to compute the growth of structure in
the IGM using combined N--body/hydrodynamics numerical simulations.
Unlike galaxy formation, the evolution of the IGM is a relatively
``clean'' problem since metal cooling is negligible and feedback
effects from star formation are arguably unimportant, or at most only
affect the ionizing radiation field. Within the context of a given
cosmology and primordial power spectrum, it is thus possible to treat the
calculation as an initial value problem, without the need for
introducing additional physical processes ``by hand.''

Several groups have performed simulations of the IGM using a variety
of techniques, and all have had remarkable success in reproducing the
principal observed properties of the Ly$\alpha$ forest (Cen et al.
1994; Zhang et al. 1995; Hernquist et al. 1996; Miralda-Escud\' e et al.
1996). An unexpected finding of these calculations is that the
absorbers arise predominantly in sheets and filaments that form
coherent structures over scales of a few megaparsecs, with a typical
thickness of a few hundred kiloparsecs. These calculations readily
reproduce the observed near power-law column density distribution for
the forest over the range $13<\log N_{\rm HI}<17$.

In this talk, I present results from a simulation of the IGM in a
Cold Dark Matter dominated cosmology (Zhang et al. 1996a,b). Standard CDM
is able to match the clustering properties of material from galaxy
cluster scales down, although some variant of the model is required to
match to the fluctuations measured by {\it COBE} on larger scales. The
simulations are performed in a flat cosmology with no cosmological
constant, and $h_{50}=1$. The normalization of the power-spectrum is
given by $\sigma_8=0.7$. Since the neutral density of individual
clouds scales like $\Omega_b h_{50}^2/ \Gamma_{\rm HI}$, the
counts of objects depend on the magnitude of the radiation field. The
results presented below are for the radiation field determined by
Haardt \& Madau (1996). The hydrodynamics is solved on an Eulerian
grid, with a comoving resolution of 75 kpc on a top grid of comoving
size 9.6 Mpc. The behavior of the gas is resolved in a second subgrid,
centered in the top grid and with a resolution 4 times greater. The
Eulerian nature of the calculation has the important advantage over
Lagrangian schemes of being able to resolve structures in underdense
regions. These structures will be shown to be crucial for interpreting
the amount of absorption by diffuse {H$\;${\small\rm I\relax}} and
especially {He$\;${\small\rm II\relax}} gas in the IGM. Further details of
the simulation are provided in Zhang (1996) and Zhang et al. (1996a, b).

\section{Baryons Regained}

Figure 1 shows the column density distribution of the clouds from the
simulation.
\begin{figure}
\plottwo{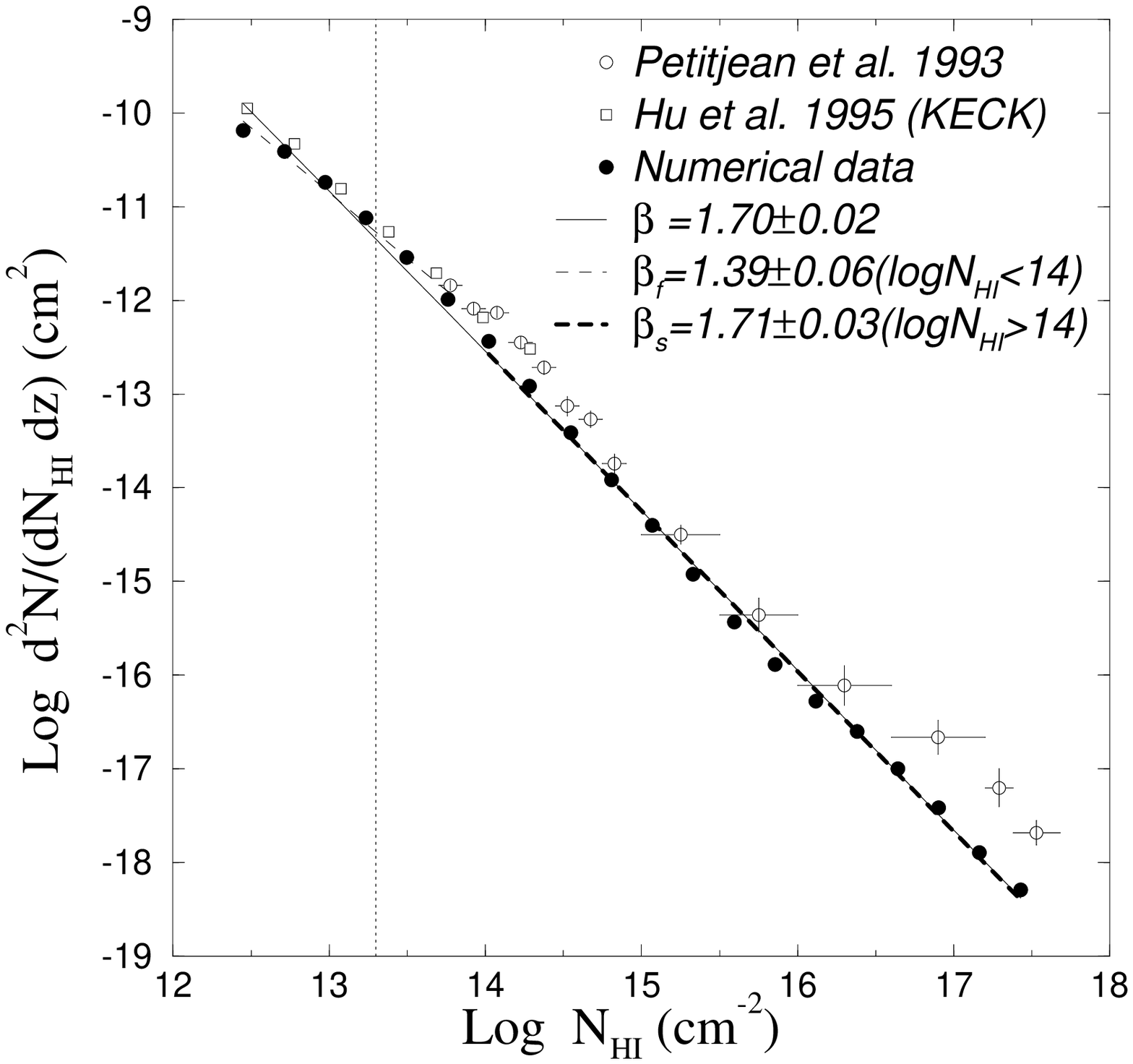}{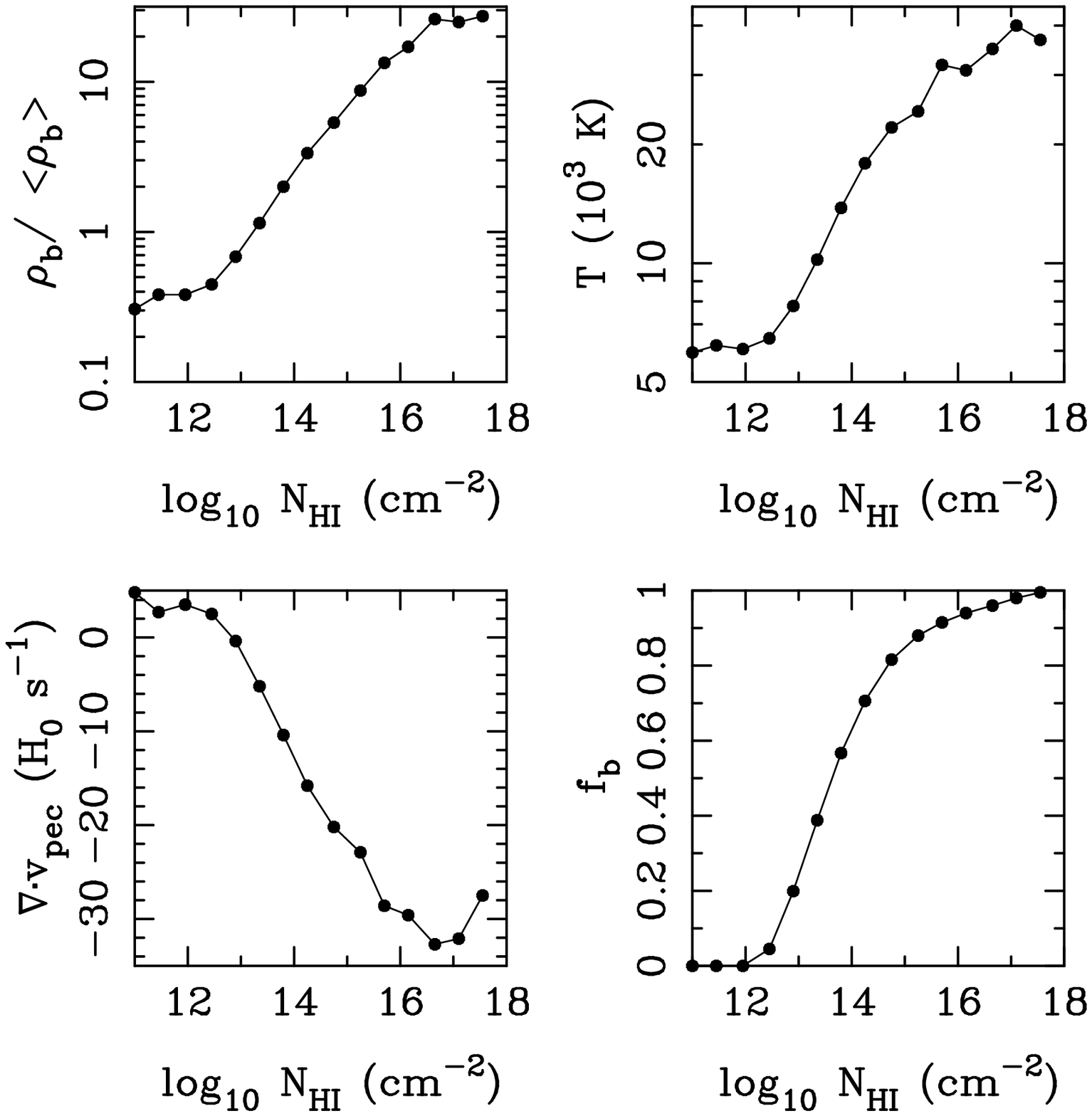}
\caption{({\it Left panel})\ The distribution of HI column
density at $z=3$. The dotted line marks the division between
optically thin and optically thick systems at line-center, for
$b=15\,{\rm km\, s^{-1}}$.
({\it Right panel})\ The mean baryon density, cloud
temperature, peculiar velocity divergence, and cumulative
baryon distribution as a function of HI column density.}
\label{fig-1}
\end{figure}
The clouds are found using a deblending algorithm for the
lines, essential to avoid undercounting weak optically thin systems
(Zhang et al. 1996a). The column densities are computed by fitting
Voigt profiles to the lines. The agreement with the observed
distribution is amazingly good, as was found by the other groups for
the optically thick systems, having {H$\;${\small\rm I\relax}}
column densities $\log N_{\rm HI}>13.3$ for a Doppler parameter of
$15\,{\rm km\,s^{-1}}$. Here, we are able to demonstrate that the
power-law behavior persists for optically thin systems as well, in
agreement with Keck measurements of the forest (Hu et al. 1995).

The simulations reveal strong correlations between the cloud physical
properties and column density (Zhang et al. 1996a,b). As shown in the
right panels of Figure 1, the higher column density systems tend to be
dense and warm, becoming more rarefied and cool toward lower values.
Most of the baryons in the universe, 60\%, are found to reside in the
optically thick absorbers. This fact alone is able to reconcile the
BBN estimate for the density of the IGM with the low Gunn-Peterson
limits and a QSO-dominated ionizing UV background, since the
Ly$\alpha$ opacity of a homogeneous component of the IGM would now be
reduced by more than a factor of 5.

The IGM, however, is found to fragment into even lower column density
systems (Hu et al. 1995). While this may not be surprising at first
glance, a comparison with the minimal column density that could result from
Jeans instability reveals a problem. A cloud with an internal density $\rho_b$
that is optically thin at the Lyman edge and in ionization equilibrium
with the metagalactic radiation field will have a neutral hydrogen density
of $n_{\rm HI}\simeq5.0\times10^{-15}\,{\rm cm^{-3}}(\rho_b/ \bar\rho_b)^2
(1+z)^6T_4^{-0.75}\Gamma_{\rm HI,-12}^{-1}$, where $\bar\rho_b$ is the cosmic
mean baryon density, here taken to correspond to $\Omega_bh_{50}^2=0.05$.
The baryonic Jeans length in a medium of mixed dark matter and baryons
is $\lambda_J=2\pi(2/3)^{1/2}c_s(1+z)/ H(z)$, where $c_s$ is the sound
speed of the baryons associated with a linear perturbation and $H(z)$ is the
Hubble constant at redshift $z$ (e.g., Bond \& Szalay 1983).
For an isothermal perturbation, $\lambda_J\simeq1\,{\rm Mpc}\,h_{50}^{-1}
T_4^{1/2}(1+z)^{-1/2}$, or about 600 kpc at $z=3$. This gives
a minimal column density due to Jeans instability of $\log N_{\rm HI}
\simeq13.5$ at $z=3$. The collapse of the system would result in an
even larger neutral column density because of the increase in the
recombination rate due to the increased density. Supposing that discrete
features may arise only from Jeans unstable systems, one might then expect
that lower column density systems cannot form. The detection of the low column
density systems becomes even more perplexing on noticing the internal cloud
densities required. Observations of the neighboring lines-of-sight
to the images of the lensed QSO candidate HE 1104--1805 (Smette et al. 1995),
place a lower limit of 100 kpc on the size of the systems with
$\log N_{\rm HI}>13.2$. For $z>3.5$, such systems will be
underdense, $\rho_b/\bar\rho_b < 1$. The existence of the optically thin
systems thus poses a puzzle.

Figure 1 shows that in the simulation, optically thin features do form,
and they indeed originate in clouds that are underdense.
The fragmentation of the IGM is complete. At most a few percent of
the baryons remain outside discrete systems to comprise a homogeneous
diffuse component. Insight into the origin of the optically thin systems
is provided by the simulation. I first discuss, however, a simple model
that offers a possible resolution to the difficulty of their occurrence.

\section{Cosmic Minivoids}

\subsection{Spherical Model}

In an open universe, the growth of linear density perturbations
``freezes out'' at an early stage, at an epoch given roughly by
$1+z_f\approx \Omega^{-1}-1$ (Peebles 1980). While bound systems will
continue to collapse, unbound systems will expand like the expansion
of the universe. In a flat universe, similar behavior will occur in a
region that is locally underdense, since any perturbation inside will
find itself in a locally open universe.

Consider now a region that is slightly underdense in an
Einstein-deSitter universe, a cosmic void. Because it is underdense,
it will grow in size slightly faster than the surrounding cosmic
expansion, increasing its underdensity. Initially ripples within
the void may grow, just as those outside, but once the void becomes
sufficiently underdense, their growth will slow. Those that are bound
will continue to collapse as they become nonlinear. But ripples that
were originally underdense, but not as strongly underdense as the
background void, will freeze (relative to the cosmic background) into
discrete structures within the void. These underdense systems would
give rise to discrete optically thin absorption features after exposure
to an ionizing radiation field, even though they are unbound.

This is illustrated in Figure 2.
\begin{figure}
\plottwo{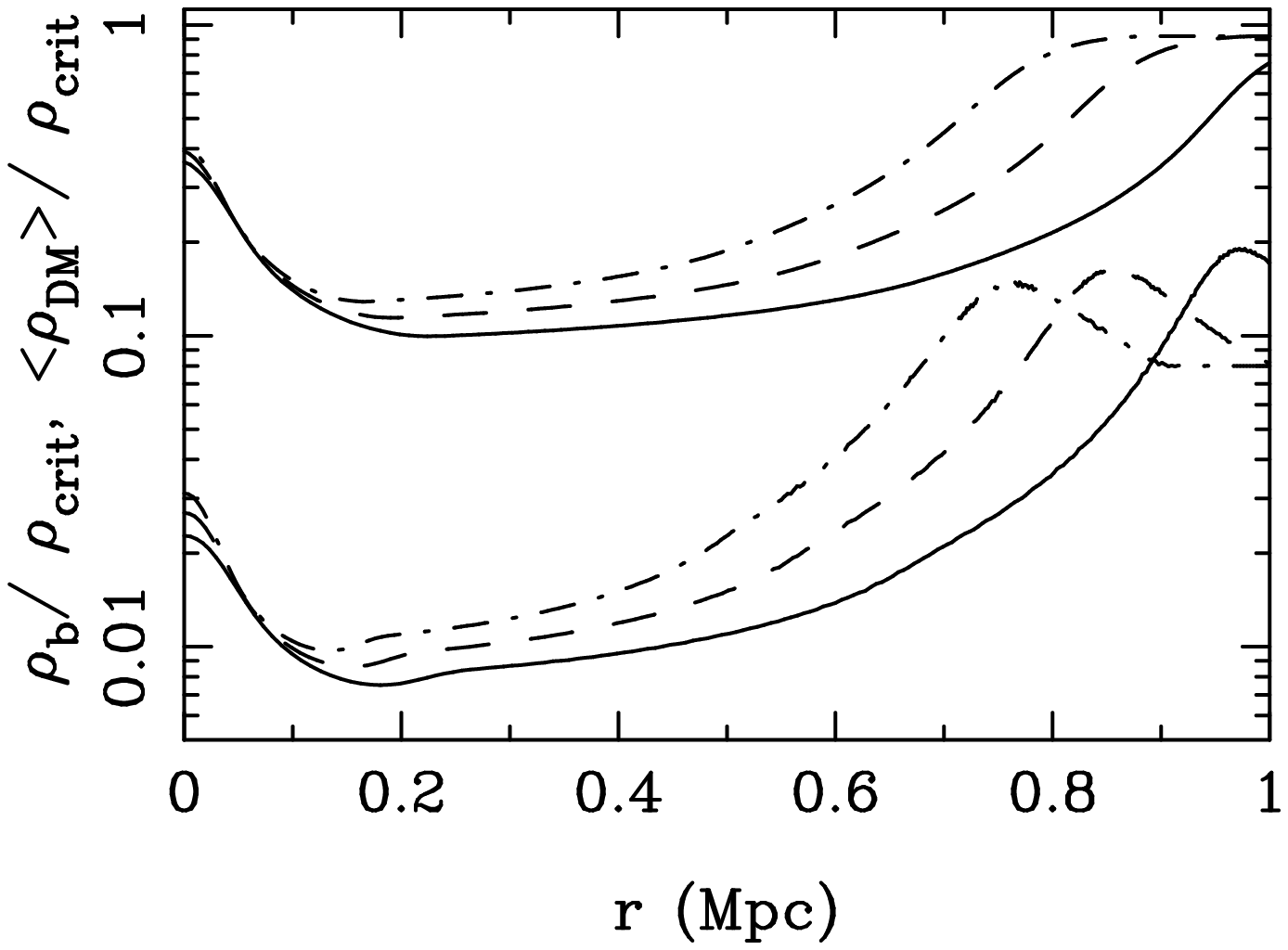}{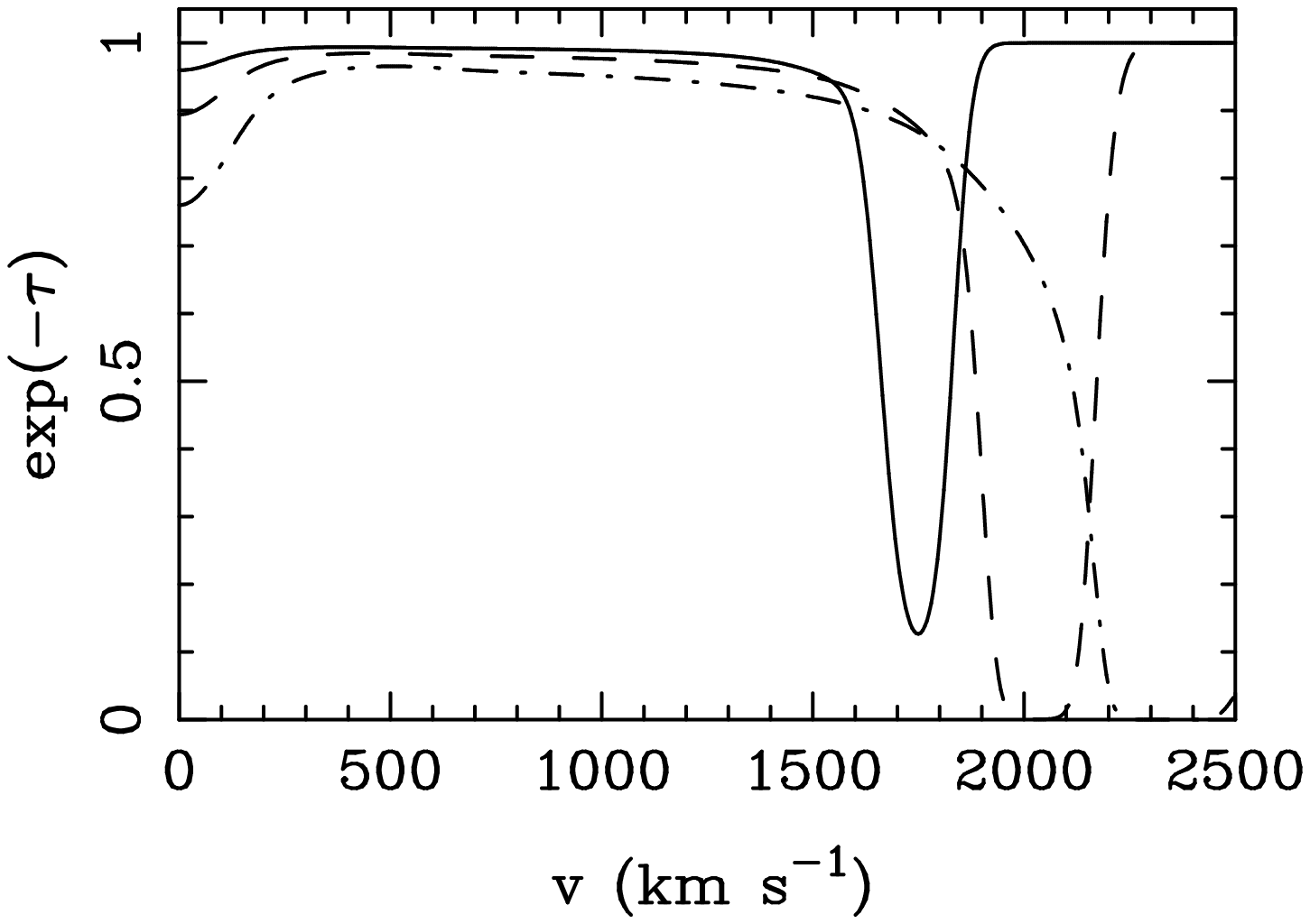}
\caption{({\it Left panel})\ The evolution of density in a spherical
minivoid. Upper curves show the dark matter density and the lower
curves show the baryon density, both normalized by the cosmic mean
density. The curves are shown at $z=3$ ({\it solid}), $z=3.5$ ({\it dashed}),
and $z=4$ ({\it dot--dashed}).
({\it Right panel})\ The spectrum for a line-of-sight passing through
the center of the minivoid. Optically thin absorption systems develop
both at the minivoid center and at its boundary. The velocity is in the
observer's frame. The solutions demonstrate that discrete absorption features
may arise from a region that is Jeans stable.}
\label{fig-2}
\end{figure}
An initially underdense compensated spherical perturbation of comoving
radius 3 Mpc is shown expanding in an Einstein-deSitter universe, with
$h_{50}=1$ and $\Omega_b=0.08$. The evolution of the dark matter and
the baryons is integrated as in Meiksin (1994). At the center of the
underdense region the density is initially raised slightly above the level of
the void (in a growing mode), but still below the cosmic average density.
A photoionizing radiation field is turned on at $z=6$, increasing slowly
with time (model ED of Meiksin \& Madau 1993).
A discrete absorption feature results along a line-of-sight passing through
the void center, with a column density of $\log N_{\rm HI}=13.1$ at $z=4$.
The perturbation is nearly frozen, with only a small decrease in the
baryon overdensity due to pressure gradients, despite the fact that the
system is Jeans stable. The reason is that the expansion velocity of the void
from its center to the edge of the feature somewhat exceeds the isothermal
sound speed of the gas. The feature is transient, but its expansion is
dominated by the flow in the void rather than by thermal pressure. A second
feature appears near the void boundary. This feature results from the
collisional nature of the gas, which is compressed near the wall of the void
where it meets the more slowly expanding external IGM. (This results in a
positive bias for the baryons just within the void boundary.)
Both mechanisms give rise to optically thin absorption features.

\subsection{Simulation Results}

\begin{figure}[t]
\plottwo{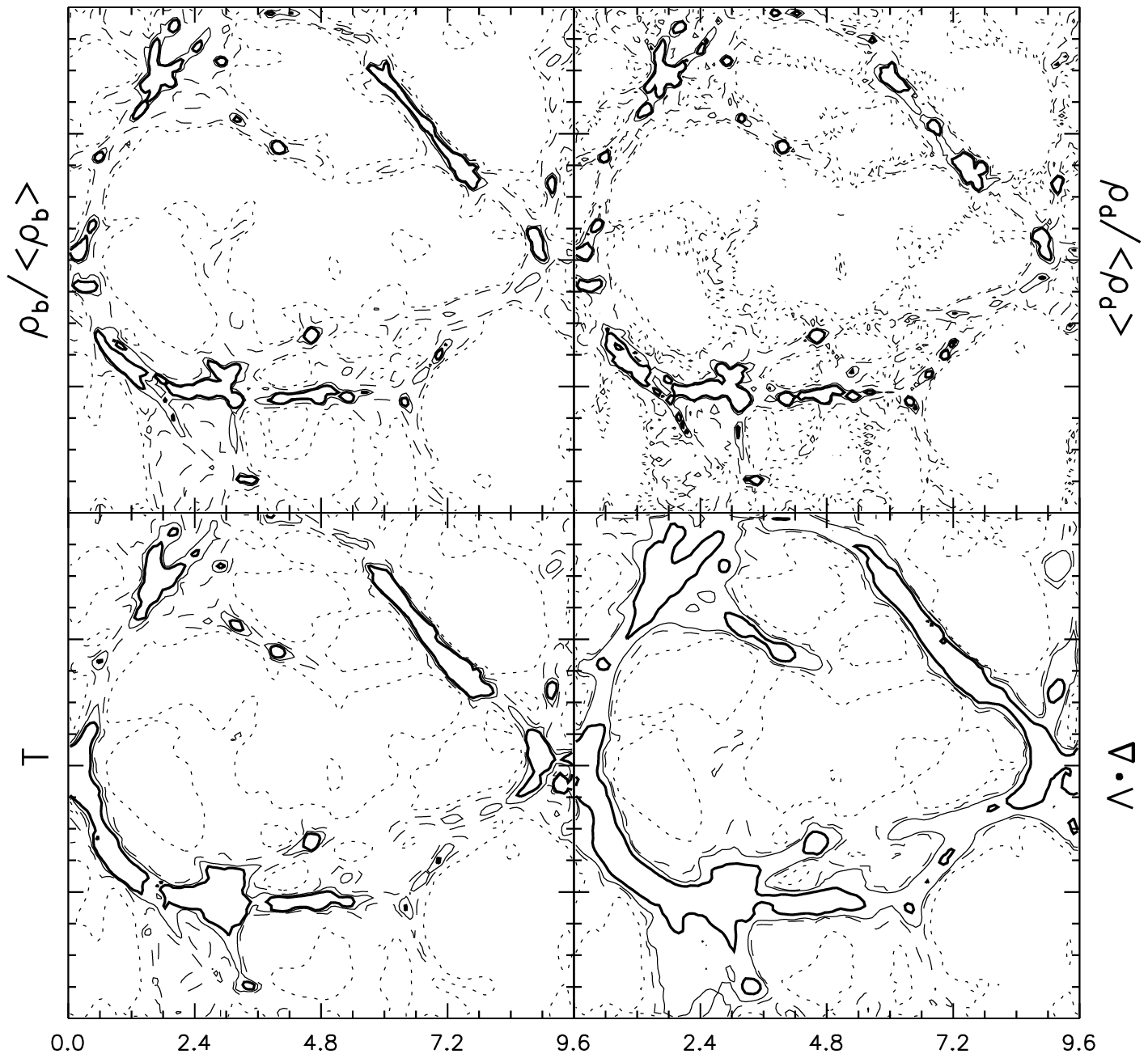}{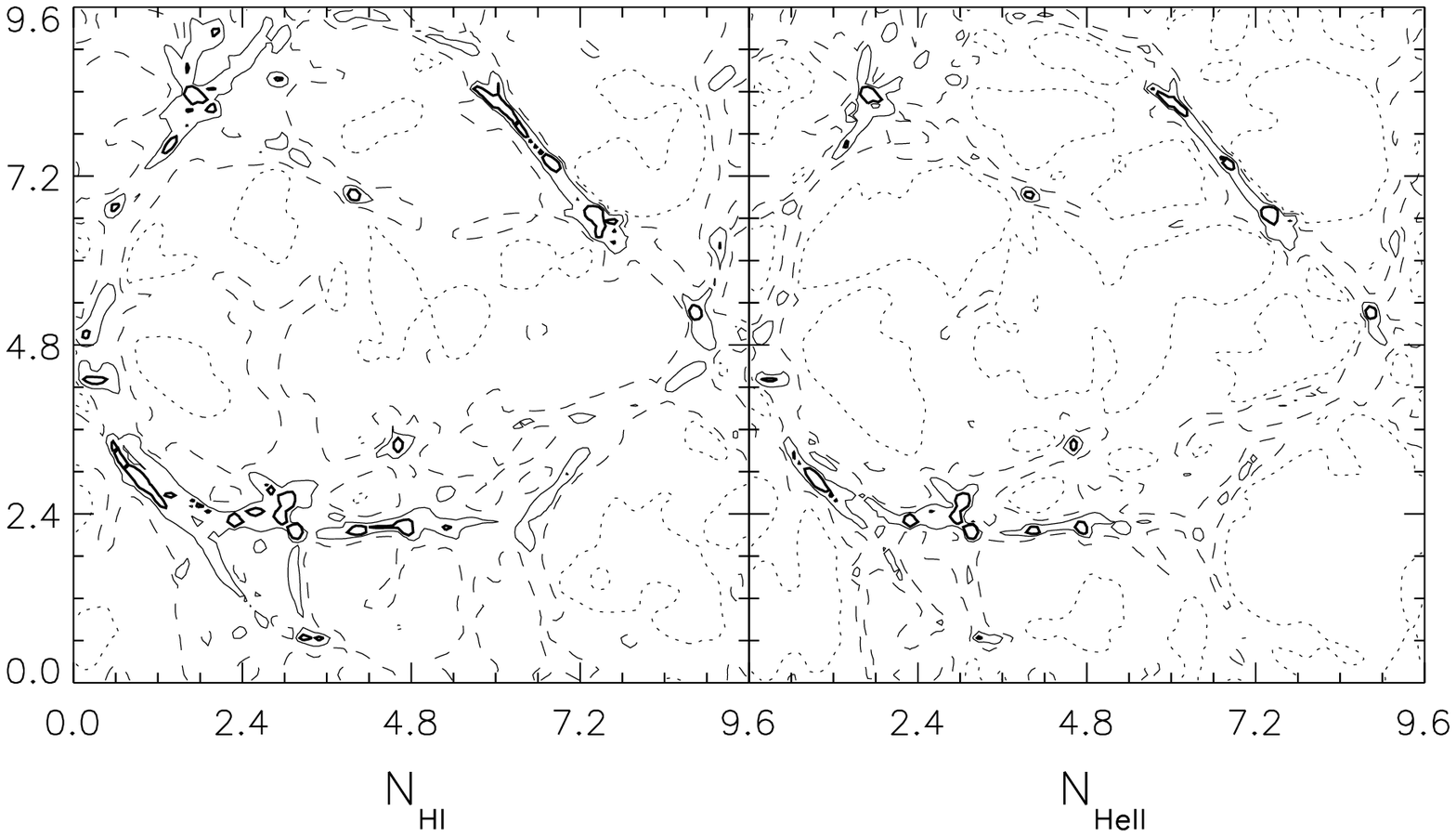}
\caption{Contour plots of a 150 kpc thick slice of the simulation box
at $z=3$, centered on a minivoid. The box size is 9.6 comoving Mpc.
({\it Left panels})\ The baryon (top left) and dark matter (top right)
contour levels are 0.5 ({\it dotted}), 1 ({\it dashed}), 3 ({\it thin solid}),
and 5 ({\it thick solid}). The temperature contours, in units of $10^3$ K,
are 6 ({\it dotted}), 10 ({\it dashed}), 14 ({\it thin solid}), and
20 ({\it thick solid}). The peculiar velocity divergence contours are,
in units of the Hubble constant at $z=3$, 5 ({\it dotted}), 0 ({\it dashed}),
-3 ({\it thin solid}), and -15 ({\it thick solid}). The low density region
bordered by the filaments exhibits density fluctuations with positive
velocity divergence. These fluctuations give rise to discrete optically
thin absorption features. ({\it Right panels})\ Contour plots of the HI
and HeII column densities.
The contour plots are at $\log N_{\rm HI}=12$ ({\it dotted}),
13 ({\it dashed}), 14 ({\it thin solid}), and 15 ({\it thick solid}), and
at $\log N_{\rm HeII}=14$ ({\it dotted}), 15 ({\it dashed}),
16 ({\it thin solid}), and 17 ({\it thick solid}) (although radiative
transfer has been ignored for HeII). The minivoid is populated by optically
thin HI systems, and saturated HeII systems.}
\label{fig-3}
\end{figure}

The numerical simulation displays elements similar to this simple
model. The underdense gas is indeed expanding faster than the cosmic
rate, as shown by its positive peculiar velocity divergence in Figure 3.
The structures form in regions a few comoving
megaparsecs across. These minivoids are bordered by the collapsed
filaments and sheets that give rise to the optically thick systems
(Figure 3). By contrast, almost all of the {He$\;${\small\rm II\relax}}
systems, even those that are saturated, form in the minivoids. Because
the effective opacity due to line-blanketing is dominated by features
that are just becoming saturated (Madau \& Meiksin 1994), the
{He$\;${\small\rm II\relax}} Ly$\alpha$ opacity in a QSO spectrum serves
as a probe of the structure of the minivoids. A more complete description
of the {He$\;${\small\rm II\relax}} absorption and the cloud structure is
provided in Zhang et al. (1996a, b).

These results admit a nontrivial consistency check based on the total
opacity of the optically thin systems. While equation (1) may not be
inverted to solve for $\Omega_D$ in the presence of clumping, a lower
limit to the cosmic baryon density may be obtained when the absorbers
are optically thin and underdense. Taking $\tau_\alpha$ to refer to
the net stochastic absorption of the optically thin systems,
$\Omega_D$ may be interpreted as the spatially averaged {\it rms}
baryon density of the optically thin systems. (This is true
only for optically thin clouds. See Jenkins \&
Ostriker 1991 for the case including optically thick absorbers.)
If the internal density of the clouds is less than the cosmic mean,
$\Omega_D$ provides a lower bound to the average cosmic baryon density.
Counting the lines in Hu et al. (1995) for which the line-center
Ly$\alpha$ opacity is less than 0.5, and doubling to account for
incompleteness (as discussed in Hu et al.), gives
$\tau_\alpha\approx0.08$ for the optically thin sytems. This yields a
lower limit for the cosmic baryon density of $\Omega_b>0.02h_{50}^{-3/2}$,
consistent with the value $\Omega_b=0.06$ adopted in the simulation.

The trends in Figure 1 permit a determination of the mass distribution
of the clouds. Over the column density range $12.5<\log N_{\rm HI}<14.5$,
the internal baryon density of the clouds scales according to
$\rho_b/\bar\rho_b\simeq N_{\rm HI,13}^{1/2}$, where $N_{\rm HI,13}$ is the
column density in units of $10^{13}\,{\rm cm^{-2}}$. The
neutral fraction scales like $f_{\rm HI}\equiv n_{\rm HI}/n_{\rm H}
\simeq3.8\times10^{-6} [(1+z)/4]^3T_4^{-0.75}\Gamma_{\rm HI,-12}^{-1}
N_{\rm HI, 13}^{1/2}$. (Note that these relations imply a roughly constant
size for the absorbers of $\sim100$ kpc over this column density range.) The
observed number of absorption systems is $\partial^2 N/ \partial
N_{\rm HI}\partial z\approx6\times10^{-13} N_{\rm HI, 13}^{-1.5}(1+z)^{2.5}$
(Hu et al. 1995). These results may be combined to derive the contribution
of the Ly$\alpha$ forest to the closure density,

\begin{equation}
\Omega_{Ly\alpha}=\frac{1.4m_{\rm H}}{\rho_{\rm crit}}\frac{H_0}{c}
\int dN_{\rm HI}\,\frac{N_{\rm HI}}{f_{\rm HI}}\frac{\partial^2 N}
{\partial N_{\rm HI}\partial z}(1+z)^{5/2}\propto
\log\left(\frac{N_{\rm HI, max}}{N_{\rm HI, min}}\right).
\end{equation}
Most of the baryons lie in the column density range $12.5<\log N_{\rm HI}
<14.5$, distributed equally per decade in column density, in accordance
with Figure 1. Only a small fraction, less than 5\%, is found to reside in
systems with column densities smaller than $10^{12.5}\,{\rm cm^{-2}}$.
Since the density of neutral hydrogen scales like the square of the baryon
density, according to this model {\it the reason it has been so difficult to
detect the Gunn-Peterson effect is that the Ly$\alpha$ opacity associated with
a homogeneous component of the IGM has been suppressed by at least two orders
of magnitude compared with the BBN value due to fragmentation}.

\section{Observational Verification}

To date, most of the evidence supporting the results of the
simulations is circumstantial. The strongest evidence is provided by
the large cloud sizes inferred from neighboring lines-of-sight to QSOs
(Smette et al. 1995; Fang et al. 1996), however the filamentary nature
of the absorbers found in the simulations has yet to be demonstrated.
The possibility that the clouds are giant spheres (though perhaps
unlikely), cannot yet be excluded. Indeed, the higher column density
systems ($\ga10^{16}\,{\rm cm^{-2}}$) may still well arise from moderate
sized spheroidal minihalos (Rees 1986; Ikeuchi 1986). (See Charlton et
al. 1996 for a discussion of the observational implications of the
simulations for neighboring line-of-sight statistics.) A strategy for
demonstrating a flattened or filamentary geometry for the absorbers is
to map their distribution on the sky using multiple lines-of-sight
with transverse separations up to a few megaparsecs. Similarly, while
the sizes and abundances of the absorbers are consistent with their
being the dominant reservoir of the baryons at high redshift ($z>2$),
the uncertainties in the cloud sizes and geometry, BBN baryon density,
and magnitude of the ionizing UV background preclude a definite
statement.

Evidence in support of the findings of the simulations may come from
an entirely different direction. Tytler et al. (1995) and Songaila \&
Cowie (1996) report the detection of {C$\;${\small\rm IV\relax}}
in the Ly$\alpha$ forest. Songaila and Cowie find a median ratio of
$N_{\rm CIV}/ N_{\rm HI}$
$\sim3\times10^{-3}$ for $15<\log N_{\rm HI}<17$. The detection of
{C$\;${\small\rm II\relax}} would permit a direct estimate of the
ionization parameter $U=n_\gamma/n_{\rm H}$, where $n_\gamma$ is the
number density of ionizing photons and $n_{\rm H}$ the total internal
hydrogen density of a cloud. The Haardt \& Madau (1996) spectrum has a
large break at the {He$\;${\small\rm II\relax}} Lyman edge, and a
spectral index of $\approx0.5$ between the hydrogen and helium Lyman
edges. Thus $n_\gamma\simeq1.8\times10^{-5}\Gamma_{\rm HI,-12}$. Using the
relation above found between cloud internal density and column
density, the ionization parameter for the clouds is $U\simeq2.3 N_{\rm
HI, 13}^{-1/2}$ at $z=3$. There is some question as to how close the
absorbers are to photoionization thermal equilibrium. Low column density
clouds will have too low a density ($n_{\rm H}<10^{-4}\,{\rm cm^{-3}}$),
to maintain photoionization thermal equilibrium, and so will be
too cool (due to the adiabatic expansion of the IGM from which they
arose), while infalling material in higher column density systems may
be shock-heated to too high a temperature (Meiksin 1994). The
temperatures are available from the simulation, and so may be used
directly (Zhang et al. 1996b). Here we give indicative values for the
metal ionization using CLOUDY (Ferland 1993), for which photoionization
thermal equilibrium is assumed. Consider a typical cloud with
a column density of $\log N_{\rm HI}=15.5$. Its internal hydrogen
density will be $n_{\rm H}\approx1\times10^{-4}\,{\rm cm^{-3}}$ and
its ionization parameter will be $U=0.13$. The neutral hydrogen fraction
in the cloud is $f_{\rm HI}\approx10^{-4.8}$, and the fractions
of {C$\;${\small\rm IV\relax}}, {C$\;${\small\rm II\relax}},
and {Si$\;${\small\rm IV\relax}} are $f_{\rm CIV}\approx10^{-1.4}$,
$f_{\rm CII}\approx10^{-3.9}$, and $f_{\rm SiIV}\approx10^{-5.2}$.
If the slabs are uniform, so that the metal column densities may be added
for comparing with the {H$\;${\small\rm I\relax}}, then
matching to $N_{\rm CIV}/N_{\rm HI}\approx0.003$ requires a
carbon abundance of about 0.4\% solar.
Both {C$\;${\small\rm II\relax}} and {Si$\;${\small\rm IV\relax}} will be
undetectable, with $N_{\rm CII}/N_{\rm CIV}\approx3\times10^{-3}$ and
$N_{\rm SiIV}/N_{\rm CIV}\approx2\times10^{-5}$ (for a solar abundance ratio
of Si to C). A detection of {C$\;${\small\rm II\relax}} matching the
predicted value, however, may be made possible
by stacking spectra showing {C$\;${\small\rm IV\relax}} absorption.
For a system with $\log N_{\rm HI}=16.5$, the situation improves. The
simulations predict $U\simeq0.04$, which gives $f_{\rm HI}\approx10^{-4.3}$,
$f_{\rm CIV}\approx10^{-0.7}$, $f_{\rm CII}\approx10^{-2.1}$, and
$f_{\rm SiIV}\approx10^{-2.9}$. The required carbon abundance is 0.2\% solar.
In this case, $N_{\rm CII}/ N_{\rm CIV}\approx0.04$, so that
{C$\;${\small\rm II\relax}} should be detectable, while
$N_{\rm SiIV}/N_{\rm CIV}\approx7\times10^{-4}$. It is critical that
measurements be performed at high resolution to ensure that the same
subcomponents of an individual Ly$\alpha$ cloud are compared.
Songaila \& Cowie (1996) report values for $N_{\rm CII}/N_{\rm CIV}$ in
good agreement with the $\log N_{\rm HI}=16.5$ prediction. Their values for
$N_{\rm SiIV}/N_{\rm CIV}$, however, are substantially higher than the
estimates provided here. This suggests that there may be a second population
of absorption systems not accounted for by the simulations
(e.g., York et al. 1986), or that the assumption of photoionization thermal
equilibrium is violated. A more careful calculation of the expected
column density ratios using the simulation results must be performed
to address this issue.

While much remains to demonstrate the correctness of the
structure of the IGM found in the numerical simulations, the ease with
which the principal observed properties of the Ly$\alpha$ forest are
reproduced in a CDM-dominated cosmology is a compelling argument in
its favor. An additional prediction of the model emphasized
here is that the Gunn-Peterson effect should be absent as a
consequence of the fragmentation of the baryons and dark matter in
cosmic minivoids. In the future, it may be possible to test rival
cosmological theories based on the properties of the IGM.

\acknowledgments
The author thanks his collaborators Peter Anninos, Mike Norman, and
especially Yu Zhang, whose numerical simulation results provided the
material for much of this talk, and to the conference organizers
for their generous hospitality.


\begin{references}
\reference Bajtlik, S., Duncan, R.~C., \& Ostriker, J.~P. 1988, \apj, 327, 570
\reference Bond, J.~R., \& Szalay, A.~S. 1983, \apj, 274, 443
\reference Cen, R., Miralda-Escud\' e, J., Ostriker, J.~P., \& Rauch, M. 1994,
\apj, 437, L9
\reference Charlton, J., Anninos, P., Zhang, Y., \& Norman, M. 1996, \apj\
(submitted)
\reference Copi, C.~J., Schramm, D.~N., \& Turner, M.~S. 1995, \apj, 455, L95
\reference Espey, B.~R. 1993, \apj, 411, L59
\reference Fall, S.~M., Pei, Y.~C. 1993, \apj, 402, 479
\reference Fang, Y., Duncan, R.~C., Crotts, A.~P.~S., \& Bechtold, J. 1996,
\apj, 462, 77
\reference Ferland, G.~J. 1993, University of Kentucky Department of Physics
and Astronomy Internal Report
\reference Field, G.~B. 1959, \apj, 129, 536
\reference Giallongo, E., et al. 1994, \apj, 425, L1
\reference Gunn, J.~E., \& Peterson, B.~A. 1965, \apj, 142, 1633
\reference Haardt, F., \& Madau, P. 1996, \apj, 461, 20
\reference Hernquist, L., Katz, N., Weinberg, D.~H., \& Miralda-Escud\' e, J.
1995, \apj, 457, L51
\reference Hu, E.~M., Kim, T.-S., Cowie, L.~L., Songaila, A., \& Rauch, M.
1995, \aj, 110, 1526
\reference Ikeuchi, S. 1986, \apss, 118, 509
\reference Jenkins, E.~B., \& Ostriker, J.~P. 1991, \apj, 376, 33
\reference Lynds, C.~R. 1971, \apj, 164, L73
\reference Madau, P., \& Meiksin, A. 1994, \apj, 433, L53
\reference Meiksin, A. 1994, \apj, 431, 109 
\reference Meiksin, A., \& Madau, P. 1993, \apj, 412, 34
\reference Miralda-Escud\'e, J., Cen, R., Ostriker, J.~P., \& Rauch, M. 1996,
\apj\ (submitted)
\reference Peebles, P.~J.~E. 1980, The Large-Scale Structure of the Universe,
Princeton: Princeton University Press, \S~11
\reference Petitjean, P., et al. 1993, \mnras, 262, 499
\reference Rees, M.~J. 1986, \mnras, 218, 25P
\reference Rugers, M., \& Hogan, C.~J. 1996, \apj, 459, L1
\reference Smette, A., Robertson, J.~G., Shaver, P.~A., Reimers, D.,
Wisotzki, L., \& Koehler, T. 1995, \aaps, 113, 199
\reference Songaila, A., \& Cowie, L.~L. 1996, \aj, 112, 335
\reference Steidel, C.~C., \& Sargent, W.~L.~W. 1987, \apj, 318, L11
\reference Tytler, D., et al. 1995, in QSO Absorption Lines, ESO Astrophysics
Symposia, G. Meylan, Heidelberg:\ Springer, 289
\reference Tytler, D., Fan, X.-M., \& Burles, S. 1996, Nature, 381, 207
\reference York, D.~G., Dopita, M., Green, R., \& Bechtold, J. 1986, \apj,
311,610
\reference Zhang, Y. 1996, Ph.D. thesis, University of Illinois at
Urbana-Champaign
\reference Zhang, Y., Anninos, P., \& Norman, M.~L. 1995, \apj, 453, L57
\reference Zhang, Y., Anninos, P., Norman, M.~L., \& Meiksin, A. 1996a,
\apj\ (in preparation)
\reference Zhang, Y., Meiksin, A., Anninos, P., \& Norman, M.~L. 1996b,
\apj\ (in preparation)
\end{references}
\end{document}